\documentclass[twoside,11pt]{article}

\usepackage{jmlr2e}
\RequirePackage{amsmath}
\RequirePackage{natbib}
\RequirePackage[colorlinks,citecolor=blue,urlcolor=blue]{hyperref}
\usepackage{graphicx,hyperref}
\usepackage{algpseudocode}
\usepackage{multirow}
\usepackage{dsfont}
\usepackage{verbatim}
\usepackage{amssymb}
\usepackage{longtable}
\usepackage{esvect}
\usepackage{mathrsfs}
\usepackage[mathscr]{eucal}
\usepackage[ruled,vlined]{algorithm2e}

\def\`#1{\textbf{#1}}
\def\*#1{\mathbf{#1}}
\def\X{\mathbf{X}}
\def\Tr{\emph{Tr}}
\def\Y{\mathbf{Y}}
\def\bLambda{\mathbf{\Lambda}}
\def\!#1{\mathcal{#1}}
\def\^#1{\left\langle{#1}\right\rangle}
\def\TPB{\mathcal{TPB}}

\def\N{\mathcal{N}}
\def\Ga{\mathcal{G}a}
\def\z_k{\langle z_k \rangle}
\def\o_k{\langle o_k \rangle}
\def\({\left(}
\def\){\right)}
\def\[{\left[}
\def\]{\right]}

\def\argmax{arg\max}

\ShortHeadings{Gene co-expression networks via biclustering}{Gao \emph{et al.}}
\firstpageno{1}

\begin{document}

\title{Differential gene co-expression networks via \\Bayesian biclustering models}

\author{\name Chuan Gao \email chuan.gao@duke.edu \\
       \addr Department of Statistical Science\\
       Duke University\\
       Durham, NC 27705, USA
       \AND
       \name Shiwen Zhao \email shiwen.zhao@duke.edu \\
       \addr Computational Biology and Bioinformatics \\
       Duke University\\
       Durham, NC 27705, USA
       \AND
       \name Ian C McDowell \email ian.mcdowell@duke.edu \\
       \addr Computational Biology and Bioinformatics \\
       Duke University\\
       Durham, NC 27705, USA
       \AND
       \name Christopher D Brown \email chrbro@mail.med.upenn.edu  \\
       \addr Department of Genetics \\
       University of Pennsylvania\\
       Philadelphia, PA 19104, USA
       \AND
       \name Barbara E Engelhardt \email bee@princeton.edu \\
       \addr Department of Computer Science \\
       Princeton University\\
       Princeton, NJ 08540, USA}


\maketitle

\begin{abstract}
Identifying latent structure in large data matrices is essential for
exploring biological processes. Here, we consider recovering gene
co-expression networks from gene expression data, where each network
encodes relationships between genes that are locally co-regulated by
shared biological mechanisms. To do this, we develop a Bayesian
statistical model for \emph{biclustering} to infer subsets of
co-regulated genes whose covariation may be observed in only a subset
of the samples. Our biclustering method, \emph{BicMix}, has desirable
properties, including allowing overcomplete representations of the
data, computational tractability, and jointly modeling unknown
confounders and biological signals. Compared with related biclustering
methods, BicMix recovers latent structure with higher precision across
diverse simulation scenarios. Further, we develop a method to recover
gene co-expression networks from the estimated sparse biclustering
matrices. 
We apply BicMix to breast
cancer gene expression data and recover a gene co-expression network
that is differential across ER+ and ER- samples.
\end{abstract}


\section{Introduction}

Cellular mechanisms by necessity tightly regulate the spatiotemporal transcription of all genes. Gene transcription is not independently regulated across genes: many of the mechanisms regulating transcription affect multiple genes simultaneously. Functional \emph{gene modules} consist of subsets of genes that share similar expression patterns and perform coordinated cellular functions~\citep{hung_2010,parkkinen_2010}. If we consider each gene as a vertex in a network, then pairs of genes within a gene module for which the correlation in expression levels cannot be explained by other genes will be connected by an undirected edge. Across all functional gene modules, these pairwise relationships constitute gene co-expression networks. Constructing these undirected gene networks, as compared to clustering genes into gene modules~\citep{eisen:1998,jiang_cluster_2004,reich_genepattern_2006,souto_clustering_2008}, provides rich detail about pairwise gene relationships. An even richer structure capturing these pairwise relationships would be a directed network of genes, but currently directed networks are computationally intractable to construct relative to undirected gene networks~\citep{friedman_2000,davidich_boolean_2008,macneil_2011,karlebach_2008}. This work describes an approach to compute an undirected gene co-expression network from a probabilistic model that clusters both genes and samples. 

Several algorithmic methods have been proposed to construct gene co-expression networks by partitioning a set of genes (and, in some cases, samples) into gene modules from which an undirected graph is elicited~\citep{zhang_2005,ruan_2010}.
In most cases, gene partitioning creates disjoint sets of genes, implying that genes only participate in a single gene module; biologically this assumption does not hold, and the impact is that the gene networks based on disjoint models are not well connected. These approaches are not probabilistic, and thus uncertainty in the clustering is not well characterized. In our work, we take a flexible statistical approach to modeling relationships in gene expression data.

\subsection{Latent factor models for gene expression data.}

Latent factor models are often used to identify groups of co-regulated genes in gene expression data~\citep{Engelhardt:2010,Carvalho:BFRM,West:2003,Bhat:2011}. In particular, latent factor models decompose a matrix $\Y\in \Re^{p \times n}$ of $p$ genes and $n$ samples into the product of two matrices, $\bLambda \in \Re^{p\times K}$, the factor loadings, and $\X \in \Re^{K\times n}$, the latent factor matrix, for $K$ latent factors, and assuming independent Gaussian noise. 
Because it is costly to obtain and assay genome-wide gene expression levels in a single sample, most gene expression studies include observations of many more genes $p$ than samples $n$. This so-called $p\gg n$ scenario limits our ability to find latent structure in this expansive but underconstrained space and suggests the use of strong regularization on the factors and loadings to provide structure for this optimization problem. For example, we may regularize a latent space to exclude a feature from contributing to all but a few latent factors through the application of encouraging sparse loading vectors~\citep{Witten:2009,Engelhardt:2010}. The resulting sparse loading matrix enables non-disjoint gene clusters to be extracted from a fitted latent factor model~\citep{Carvalho:BFRM}. Sparse latent factor models are much more interpretable than their non-sparse counterparts, but are generally limited to finding a small number $K$ of large clusters for reasons that include computational tractability and robustness. 

Given the complexity of biological systems, manifested as tens of thousands of genes, subsets of which are transcriptionally regulated by possibly a larger number of mechanisms, a more appropriate model might create a large number $K$ of gene clusters that each contain small numbers of genes. Constructing an \emph{overcomplete} representation of the data, where the number of latent factors is larger than the number of samples, has proven effective for a variety of applications in classical statistics~\citep{Elad:2006KSVD,Mairal:2010,Witten:2009,Zou2009,Xing2012}. However, building such a method is difficult because, without careful regularization, the recovered factors are not robust or interpretable, and signals that account for a small proportion of the variation in the gene expression data are not reproducibly recovered.

Besides encouraging sparsity in the factor loading matrix, which results in non-disjoint clusters of genes that co-vary across all samples, one can also induce sparsity in the factor matrix, which results in non-disjoint subsets of samples within which small number of genes uniquely exhibit co-variation. Mathematically, this corresponds to regularizing both factor and loading matrices using priors that induce zero-valued elements.
Biologically, biclustering model components identify covariation among a small number of genes that is exclusive to, for example, samples from adipose tissue. This approach addresses a general problem known as biclustering~\citep{cheng2000,ben-dor2003,murali2003,li_qubic2009,prelic2006,bergmann2003,huttenhower2009,lazzeroni_plaid2000,gu_bayesian2008,bozdag2009,hochreiter2010,kluger2003,aguilar-ruiz2005}. A biclustering model decomposes a data matrix into clusters that each correspond to a subset of samples and a subset of features that exhibit latent structure. Our flexible Bayesian approach allows each sample and each gene to belong to any number of the $K$ latent clusters or \emph{components} (i.e., a sparse loading vector of length $p$ and a sparse factor vector of length $n$), and does not require orthogonality across the factors or loadings.

\subsection{Capturing independent sources of variation in gene expression data.}

Gene expression levels have been shown to be sensitive to a number of environmental and technical covariates such as batch, sex, ethnicity, smoking status, or sample tissue heterogeneity. 
Methods to adjust the observation matrix to remove the effects of these covariates without eliminating signals of interest have been proposed, but most attempts have been limited to correcting for confounding effects in a preprocessing step~\citep{Pickrell:2010,Brown:2013} or correcting for confounding effects jointly with univariate association testing~\citep{Leek2007,Stegle2010,Listgarten:2010}. The two-stage approach applied to estimates of co-expression network have not been successful: often variation in expression levels of large gene modules are captured in the confounding effects and controlled in the first step. 
Because of the exploratory nature of this method, we will not include association testing in our models, but instead develop factor analysis-based methods to recover all of the sources of co-variation in the observed data matrix in an unsupervised way~\citep{Gao2013}.
To recover gene co-expression signals in the presence of large-scale co-variation due to confounders, we have found that two properties of the statistical model are essential: i) co-regulated genes and confounding effects need to be modeled jointly, and ii) the number of gene clusters and the number of confounding effects must be estimated from, and scale with, the observed data~\citep{Gao2013}. 

In this paper, we develop Bayesian statistical model for biclustering
called \emph{BicMix}. Our motivation behind developing this method was
to identify large numbers of subsets of co-regulated genes capturing
as many unique sources of gene transcription variation as possible. We
next developed a simple method to reconstruct gene co-expression
networks based on the sparse covariance matrices reconstructed using
our biclustering model. This method recovers different types of gene
co-expression networks, categorized by quantifying the contribution of
each sample to the gene cluster: i) ubiquitously-expressed
co-expression networks, ii) co-expression networks specific to a
sample subtype, and iii) networks that are differentially co-expressed
across sample subtypes.  We apply this approach to two gene expression
data sets without correcting for confounding effects in the gene
expression levels. 
We apply our biclustering model to gene expression levels measured in
heterogeneous breast cancer tissue
samples~\citep{veer2002,van_de_vijver_2002} to recover a co-expression
network that is differentially expressed across ER+ and ER- samples.

\section{Results}

\subsection{Bayesian Biclustering using BicMix.}

Biclustering was first introduced to detect clusters of states and years that showed similar voting patterns among Republicans in national elections~\citep{hartigan1972} and was later referred to as \emph{biclustering} to identify similarly expressed genes~\citep{cheng2000}. It has also been referred to as two mode clustering~\citep{van2004}, subspace clustering~\citep{patrikainen2006,kriegel2009}, or co-clustering~\citep{yoon2007} in various applied contexts. Biclustering was used successfully to explore latent sparse structure in different applied domains~\citep{busygin2008}, including gene expression data~\citep{cheng2000,madeira2004,madeira2009,ben-dor2003,turner2005,santamaria2007}, neuroscience~\citep{neng2009}, time series data~\citep{madeira2009}, and collaborative filtering~\citep{de_castro2007}. There are a few comprehensive reviews of biclustering for further details~\citep{prelic2006,eren2012}.

Biclustering approaches fall into four general categories. The first category of biclustering assumes that each gene is a linear combination of a mean effect, a column effect, and a row effect, some of which may be zero~\citep{cheng2000}. One approach in this category, \emph{plaid}, assumes that the gene effects are the sum of many sparse submatrix components, where each submatrix includes non-zero values only for a small number of genes and a small number of samples~\citep{lazzeroni_plaid2000,gu_bayesian2008}. 
The second category of biclustering method explicitly identifies similar samples and features in the data matrix and groups them together through hierarchical clustering~\citep{eisen:1998}. For example, samples may be clustered by considering some measure of feature similarity~\citep{ben-dor2003,bergmann2003,murali2003,aguilar-ruiz2005,bozdag2009}. The third category of biclustering method builds up biclusters by iteratively grouping features in a greedy way (e.g., identifying all genes that have correlated expression levels with a selected gene) and then removing samples that do not support that grouping~\citep{li_qubic2009}.
The last category of biclustering method uses Bayesian sparse factor analysis models~\citep{hochreiter2010}. These models decompose a gene expression matrix into two sparse matrices. Sparsity-inducing priors, such as the Laplace prior, are imposed on both the loading and the factor matrices to produce zero-valued elements in the two latent matrices. Our approach falls into this last category of a sparse statistical model for biclustering.

Specifically, we developed a Bayesian biclustering model, \emph{BicMix}, built on factor analysis with sparsity-inducing priors on both of the low dimensional matrices. In particular, we define the following latent factor model for matrix $\*Y \in \Re^{p \times n}$, which is the set of observations for $p$ genes across $n$ samples:
\begin{align}
    \*{Y=\Lambda X} + \boldsymbol{\epsilon}
\end{align}
where $\*\Lambda \in \Re^{p \times K}$ is the \emph{loading matrix},  $\*X \in \Re^{K \times n}$ is the \emph{factor matrix}, $\*\epsilon \in \Re^{p \times n}$ is the residual error matrix, and $K$ is fixed a priori. We assume that the residual error is independent across genes and samples and has a zero-mean multivariate Gaussian distribution with gene-specific variance: $\epsilon_{\cdot,i} \sim \mathcal{N}_p(0,\*\Psi)$ for $i = 1,\dots, n$, where $\*\Psi = \text{diag}(\psi_1,\dots,\psi_p)$. While a value must be given for $K$, the number of latent factors, the model removes factors when appropriate, so $K$ should be an overestimate of the number of latent factors.

To induce sparsity in both the factors and the loadings, we used the three parameter beta ($\TPB$) distribution~\citep{Armagan2011}, which has been shown to be computationally efficient and to induce flexible modeling behavior as a sparsity-inducing prior. In previous work~\citep{Gao2012,Gao2013}, we included three layers of regularization via the $\TPB$ distribution to induce sparsity in the loadings matrix; we extended this model to include this same sparsity-inducing prior on the factor matrix (see Appendix A). With this prior on both the factor and loadings matrices, the model becomes a biclustering model, estimating subsets of genes for which correlation is only observed in a subset of samples. This structured prior produces favorable properties in this biclustering model: i) the number of factors and loadings are determined by the data, because the sparsity-inducing prior removes unnecessary factors; ii) each factor and corresponding loading has a different level of shrinkage applied to it, enabling a non-uniform level of sparsity and corresponding percentage of variance explained (PVE) for each factor and loading pair~\citep{Gao2013}; iii) neither the clusters of genes nor the clusters of samples are disjoint, so all genes and all samples may be in any number of clusters, or none; and iv) strong regularization allows overcomplete estimates of the response matrix, with possibly more factors $K$ than samples $n$.

\begin{figure}
\begin{center}
\includegraphics[width=0.93\textwidth]{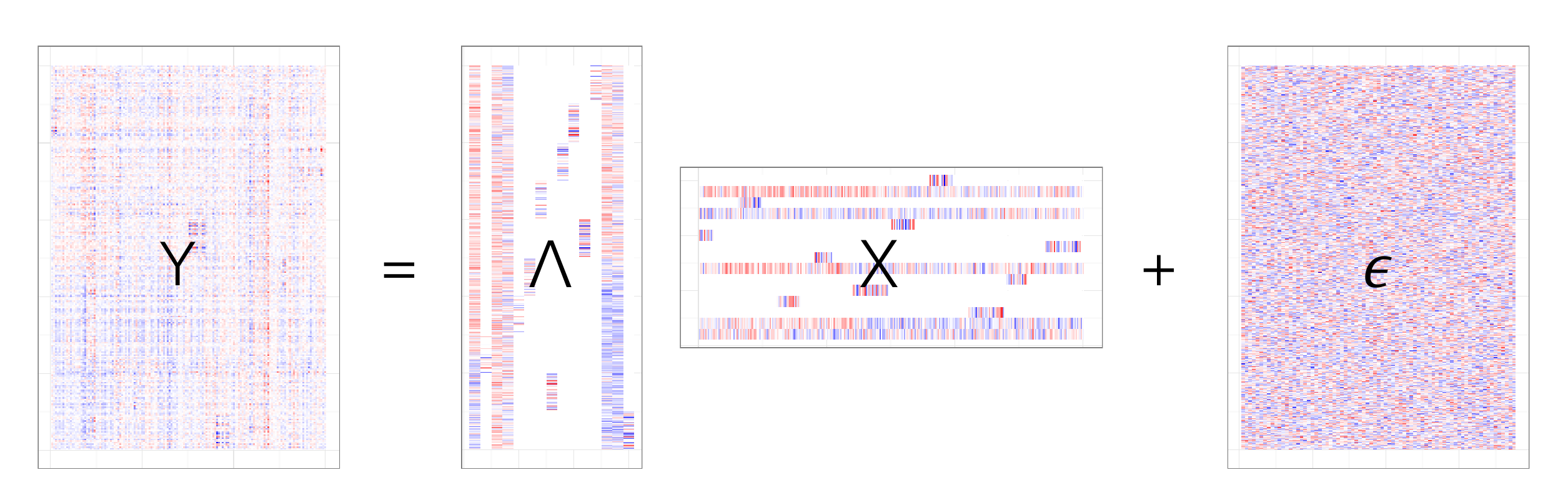}
\end{center}
\vspace{-10mm}
\caption{{\bf Schematic representation of the BicMix biclustering model.} Ordered from left to right are, respectively, the $p \times n$ gene expression matrix $\Y$, the $p \times K$ loading matrix $\Lambda$ including both sparse and dense columns, the $K \times n$ factor matrix $\X$ including both sparse and dense rows, and the $p \times n$ residual error $\epsilon$. Blue, red and white entries in each matrix correspond to negative, positive, and zero values, respectively. In the response matrix $\Y$, the $p$ genes are on the rows and $n$ samples are on the columns.}
\label{fig:illustration}
\end{figure}

In gene expression data, observed covariates or unobserved confounding effects may systematically influence variation in the observation~\citep{Leek2007,Listgarten:2010,Stegle2010}. As in prior work, we tailored our sparsity-inducing prior to jointly model these often dense confounding effects~\citep{Gao2013}. In particular, we adapted our model so that the loadings and factors are drawn from a two-component mixture distribution, where each vector is either sparse---with many zeros---or dense---with no zeros (Figure~\ref{fig:illustration}). We extract information about whether a vector is sparse or dense directly from the fitted model parameters using the expected value of the mixture component assignment variables for each component $k=1,\dots,K$, where $z_k \in \{0,1\}$ indicates a dense or a sparse loading and $o_k \in \{0,1\}$ indicates a dense or a sparse factor. This two-component mixture in the prior distribution for the factors and loadings adds two additional favorable properties to the biclustering model. First, it jointly models covariates that regulate variation in most genes and also in few genes; we have found that confounding effects are often captured in the dense components as large numbers of samples and genes are affected (e.g., batch effects)~\citep{Leek2010,Gao2013}.
 Second, it has the effect of relaxing a computationally intractable space, enabling robust and scalable parameter estimation in a Bayesian framework. Specifically, considering all possible subsets of genes and samples to identify biclusters is intractably difficult; however, it is computationally tractable to first search over the space for which cluster membership is a continuous value and then subsequently identify clusters by iteratively shrinking to zero elements with membership values near zero. We estimate parameters in this model using both Markov chain Monte Carlo (MCMC) approaches and a variational Expectation-Maximization (EM) approach; see Appendix A and B for details.

\subsection{Simulations and comparisons.}

We simulated data from an alternative generative model for the observations matrix $\* {Y=\Lambda X + \epsilon}$, where $\*Y$ has dimension $p=500$ by $n=300$ and $\epsilon_{i,j} \sim \N(0, \nu)$. Within this model, we simulated sparsity as follows: for each loading and factor, a number $m \in [5, 20]$ of elements were randomly selected and assigned values drawn from $\!N(0,2)$; the remaining elements were set to zero. We allowed components to share as many as five elements. Simulation 1 (Sim1) has ten sparse components. Simulation 2 (Sim2) has ten sparse components and five dense components, for which the loadings and factors were drawn from a $\!N(0,2)$ distribution. The components were shuffled so that a sparse loading may correspond to a dense factor, and vice versa. For both simulations we considered low and high noise scenarios, so the residual error simulation parameter in the low noise (LN) setting was $\nu = 1$ and the high noise (HN) setting was $\nu = 2$. Each simulation was repeated ten times. 

We ran BicMix and five other biclustering methods---Fabia~\citep{hochreiter2010}, Plaid~\citep{lazzeroni_plaid2000}, CC~\citep{cheng2000}, Bimax~\citep{prelic2006}, and Spectral biclustering~\citep{kluger2003}---on the simulated data. For all simulations, we ran BicMix by setting $a=b=c=d=e=f=0.5$ and $\nu=\xi=1$ to recapitulate the horseshoe prior at all levels of the hierarchy. The algorithm was initialized with warm start values by running MCMC for 100 iterations and using the final sample as the initial state for variational EM. For BicMix results, components that are classified as sparse are not thresholded post hoc, because our parameter estimation tends to compute zero for these values. 
All other methods were run using their recommended settings (details in Appendix C).
For Sim2, we corrected the simulated data for the dense components by controlling for five PCs before all other methods were run; without this initial correction for dense components, results from the other biclustering methods were uninterpretable. For all runs, BicMix was initialized with $K=50$ latent factors; all other methods were initialized with the correct number of sparse factors $K=10$. For Fabia, we ran the software in two different ways. The results from running Fabia with the recommended settings are denoted as \emph{Fabia}. We also set the sparsity threshold in Fabia to the number (from $100$ quantiles of the uniform distribution over $[0.1,5]$) that produced the closest match in the recovered matrices to the number of non-zero elements in the simulated data; we label these results \emph{Fabia-truth}.

We used the \emph{recovery and relevance score} (R\&R score)~\citep{prelic2006} to measure the power and accuracy of each method in recovering true biclusterings. Let the true set of sparse matrices be $\*M_1$ and the recovered set of sparse matrices be $\*M_2$; then the R\&R score is calculated as:
\begin{align}
\text{Rec}=\frac{1}{|M_1|}\sum_{b_1 \in M_1}\max_{b_2 \in M_2}\frac{b_1 \cap b_2}{b_1 \cup b_2},\\
\text{Rel}=\frac{1}{|M_2|}\sum_{b_2 \in M_2}\max_{b_1 \in M_1}\frac{b_1 \cap b_2}{b_1 \cup b_2}.
\end{align}
\emph{Recovery} quantifies the proportion of true clusters that are recovered (i.e., recall); \emph{relevance} refers to the proportion of true clusters identified in the recovered clusters (i.e., precision). For BicMix, we applied this R\&R score to the components for which both the loading and the factor were sparse, which indicates a biclustering.
For the doubly-sparse latent factor models, Fabia, and BicMix, we also calculated a sparse stability index (SSI)~\citep{Gao2013} to compare the recovered and true matrices; SSI is invariant to label switching and scale.

For Sim 1, we found that BicMix recovered the sparse loadings, sparse factors, and the biclustering well in the low noise scenario based on both RR (Figure 2a) and SSI (Figure 2b). Fabia had the second best performance based on RR and SSI. 
For comparison, Fabia-truth achieved better R\&R scores than Fabia (Figure 2a); the clustering results from BicMix dominated those from Fabia-truth, although there was only a small gain in relevance in the low noise Sim1 results for BicMix.
Plaid showed high relevance for the recovered biclusters regardless of the noise level for Sim1, but at the expense of poor recovery scores. The remaining methods did not perform well in these simulations with respect to the R\&R score.

For Sim2, BicMix correctly identified the sparse and dense components  (Figure 2a), where a threshold of $\^{z_k} > 0.9$ was used to determine when a loading $k$ was dense.
The performance of Fabia on Sim2 deteriorated substantially relative to its performance on Sim1, although the confounders were removed from the data using the correct number of PCs and the correct number of factors was given. 
For both BicMix and Fabia, additional noise in the simulation made bicluster recovery more difficult, as shown in deterioration of the recovery score for both methods; however, unlike Fabia, the relevance score of the biclustering from BicMix was unaffected by additional noise in Sim2-HN. 
The other methods show inferior performance relative to BicMix and Fabia on this simulation. CC assumes that genes in each bicluster have constant expression values, which limits its ability to cluster genes with heterogeneous expression levels.
Bimax assumes binary gene expression values (i.e., over- or under-expressed), which also limit its utility for heterogeneous expression levels. Spectral biclustering methods impose orthogonal constraints on the biclusters; this orthogonality assumption is certainly violated in the simulations we designed here and also in gene expression data.

\begin{figure}
\begin{center}
\includegraphics[width=0.9\textwidth]{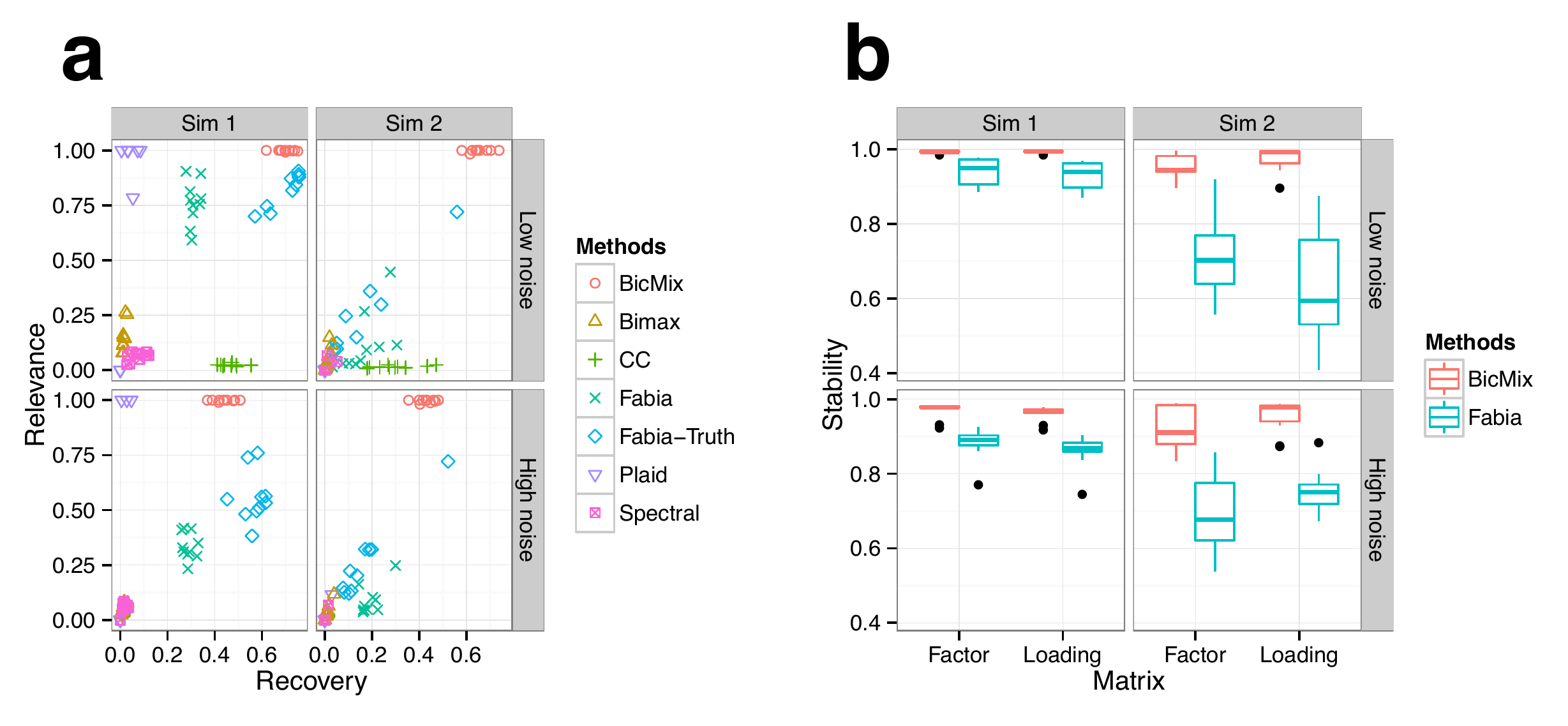}
\end{center}
\vspace{-10mm}
\caption{{\bf Comparison of BicMix with related methods.} Top row: simulation with low noise, bottom row: simulation with high noise. Left column: Sim1 with only sparse components, right column: Sim2 with sparse and dense components.  Panel a: Recovery score on the x-axis, relevance score on the y-axis for all methods in the legend. Panel b: Stability statistic (y-axis) for the sparse components recovered by BicMix and Fabia.}
\label{fig:Comparison}
\end{figure}

\subsubsection*{Gene co-expression networks from biclusters.}
To construct an undirected gene network, we built a Gaussian Markov
random field, or a Gaussian graphical model
(GGM)~\citep{schafer_empirical_2005}, using the components recovered
from our biclustering model (Appendix D). In particular, regularized
estimates of the gene-by-gene covariance matrix may be computed from
our parameter estimates; factor analysis is often viewed as a method
for low-rank covariance estimation by marginalizing over the factors,
$\X$.  Furthermore, for any subset of components with sparse loading
vectors, $A \subseteq \{1,\dots,K\}$, $\*{\Omega_A = \Lambda_A
  \Sigma_{A,A} \Lambda_A^T+\Psi}$, where $\*\Sigma_{A,A}$ is the
covariance matrix for $\*X_A$, estimates a regularized covariance
matrix for the genes loaded on $\*\Lambda_A$. Note that $\*\Omega_A$
is both sparse and full rank; biclustering is a highly structured
approach to estimating regularized covariance
matrices~\citep{schafer_shrinkage_2005}. The inverted covariance
matrix is a symmetric precision matrix $\*{\Delta_A = \Omega_A^{-1}}$,
where $\Delta_{i,j}$ quantifies the pairwise partial correlation
between genes $i$ and $j$. The intuition is that, in a GGM, edges are
defined as pairs of nodes for which the partial correlation is
non-zero. Since each loading $\Lambda_i$, $i \in A$, specifies a
cluster of genes, we do not invert the full covariance matrix, but
instead invert the submatrix that corresponds to genes with non-zero
loadings in those components. This approach avoids inducing edges
between genes that never occur in the same cluster. We used
GeneNet~\citep{schafer_empirical_2005} to test the precision matrix
for significant edges. GeneNet assumes that the edges are drawn from a
mixture of the null (no edge) and alternative (edge) distributions,
$f(E)=\eta_0 f_0(E) + \eta_A f_A(E)$ to calculate the probability of
each edge being present or not. Practically, we selected edges with a
probability of $>0.8$.

To recover co-variance networks with particular semantics with respect
to the samples, we choose the subset of components that contribute to
this covariance matrix carefully. In particular, when we select subset
$A$ to include only components that have non-zero factors in a single
tissue, we identify tissue-specific components. When we select $A$
such that all samples have a non-zero contribution to a component, we
recover ubiquitous components. And when we select $A$ such that the
mean ranks of the values for one covariate type is different than the
mean ranks of values for a different covariate type, we identify
components that are differential across the two covariate types.

\subsection{Breast cancer network}

We investigated a breast cancer data set that contains 24,158 genes
assayed in 337 breast tumor
samples~\citep{van_t_veer:2002,van_de_vijver_2002,NKI} after filtering
for genes that are $>10\%$ missing and imputing the rest of the
missing values~\citep{hastie-imputing:1999} (Appendix C).  All
patients in this data set had stage I or II breast cancer and were
younger than 62 years old. Among the 337 patients, 193 had lymph-node
negative disease and 144 had lymph-node positive disease; prognostic
signatures such as \emph{BRCA1} mutations, Estrogen Receptor (ER),
Distant Metastasis Free Survival (DMFS) were also collected for all
patients. We focused on building differential gene co-expression
networks across ER positive (ER+) and ER negative (ER-) patients
because of ER's prognostic value in profiling breast cancer
patients~\citep{Zhang-ER:2014}: cancer patients that are ER+ are more
likely to respond to endocrine therapies than patients that are
ER-. In these data, there are 249 ER+ and 88 ER- patients as compared
to 97 \emph{BRCA1} mutation carriers, two \emph{BRCA2} mutation
carriers and 97 patients with no \emph{BRCA} mutations.

\begin{figure}
\begin{center}
\includegraphics[width=0.9\textwidth]{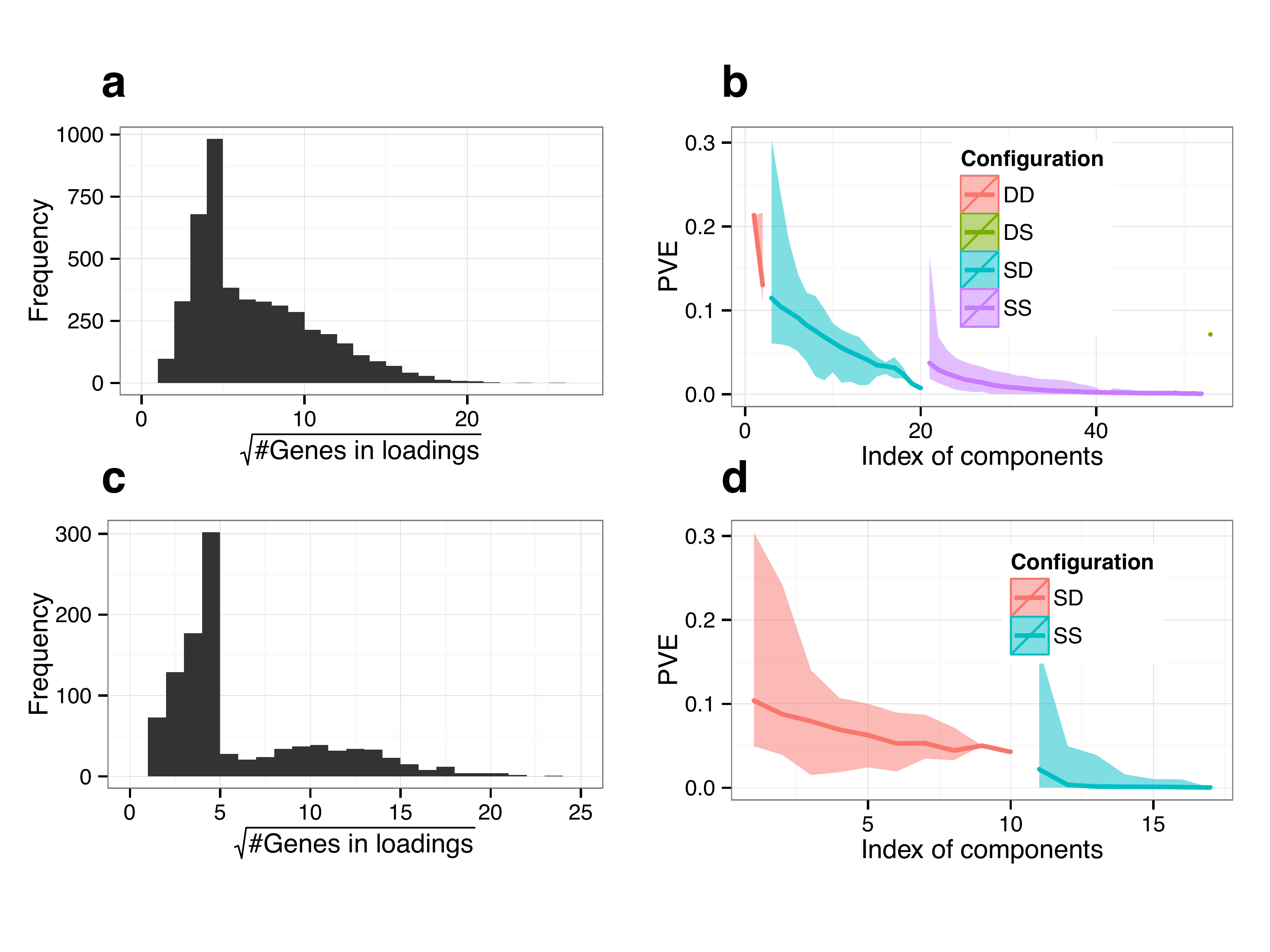}
\end{center}
\vspace{-10mm}
\caption{{\bf Distribution of the number of genes and PVE in the breast cancer data.} Panel a: Distribution of the number of genes for the 4,751 components. Panel b: PVE for two DD components, 50 SD components, 50 SS components and one DS component. Panel c: Distribution of the number of genes for the 1,036 components that are ER- specific, ER+ specific, and ER differential. Panel d: PVE for all components that are ER-, ER+, and ER differential (SS and SD only). For panels b and d, the middle lines show the median PVE, the ribbons show the range of the minimum and maximum value of PVE across 177 runs.
}
\label{fig:PVE-breast}
\end{figure}

We ran \emph{BicMix} on these data, setting $a=b=c=d=e=f=0.5$ and
$\nu=\xi=1$ to recapitulate the horseshoe prior, as in the
simulations; the initial number of components was set to $K=300$.
BicMix was run for $5,000$ iterations starting from $177$ random
values.  We first removed factors that were zero; then we removed
components for which the number of genes in the loading and the number
of samples in each factor had changed in the most recent $2,000$
iterations of EM, indicating lack of stability.  We recovered $4,721$
components across $177$ runs, of which $9$ loadings and $1,632$
factors were dense (Figure~\ref{fig:PVE-breast}a,b). 

The distribution of the number of genes in each sparse component was
skewed to small numbers (Figure~\ref{fig:PVE-breast}a). We categorized
each component as one of four types of configurations: sparse gene
loadings with sparse sample factors (SS), sparse gene loadings with
dense sample factors (SD), dense loadings with sparse factors (DS),
and dense loadings with dense factors (DD). SS components will capture
subsets of genes that are uniquely co-expressed in a subset of samples
(e.g., ER+ specific interactions). SD components capture subsets of
genes that are differentially co-expressed among all samples (e.g.,
sex-differential networks, batch effects). DS components capture a
subset of samples in which all genes have additional co-variation
(e.g., sample contamination). DD components capture variation that
occurs across all samples and affects co-variation across all genes
(e.g., latent population structure).

For each random run, we calculated the percentage of variance
explained (PVE) per component as
$\frac{\Tr\(\Lambda_i\^{X_iX_i^T}\Lambda_i^T\)}{\Tr\(\*\Lambda\^{\*{XX^T}}\*{\Lambda^T}\)}$,
where $\Tr$ denotes the trace operator. We ordered the components by
PVE within each SS, SD, DS, and DD component category. We calculated
the mean, maximum, and minimum values of the PVE-ordered, categorized
components across the random runs. Note that, because there are no
orthogonality constraints, it is possible that many of these
components explain similar variation in the observations; for this
quantification we are assuming this PVE is disjoint and normalizing
across all component-wise PVEs. We also calculated the total PVE
explained by each component category by summing the total PVE for all
components jointly in each category.  The distribution of the number
of genes contained in each loading and the PVE by component and across
SD, SS, DS, DD categories show (Figure 5). The number of components
that fell into the SS, SD, DS, DD categories accounted for $65.3\%$,
$34.5\%$, $0.085\%$, $0.1\%$, respectively, of the total number of
components. In the same order, components in the four categories
accounted for $23.3\%$, $76\%$, $0.5\%$ and $0.2\%$ of the total PVE.

We selected components from the fitted BicMix model to identify ER+
and ER- specific gene clusters (Appendix D). Moreover, to select gene
clusters that were differentially expressed across ER+ and ER-
samples, we identified components corresponding to factors that had a
significant difference in the mean rank of the factor value between
the ER+ and ER- samples based on a Wilcoxon rank sum test (threshold
$p\leq 1\times 10^{-10}$).  Across our components, we found $41$
components unique to ER+ samples, $183$ components unique to ER-
samples, and $812$ components that were differential across ER+ and
ER- samples (Figure~\ref{fig:PVE-breast}c,d).

The precision matrices of the subsets of components corresponding to the three network subtypes were constructed and edges among these genes were tested using our method for extracting gene co-expression networks from the fitted biclustering model (Appendix D)~\citep{schafer_empirical_2005}. For the ER- specific network, we recovered a total of $216$ genes and $728$ edges; for the ER+ specific network, we recovered $932$ genes and $5,383$ edges (Supplemental Figures S2-S3; Supplemental Tables S13-14). There were no replicated edges for the ER+ samples and ten edges that were replicated two or more times. We confirmed that there were no nodes and edges shared across the two specific networks. There were ten genes that were shared across the two specific networks. 

\begin{figure}
\begin{center}
\includegraphics[width=1\textwidth]{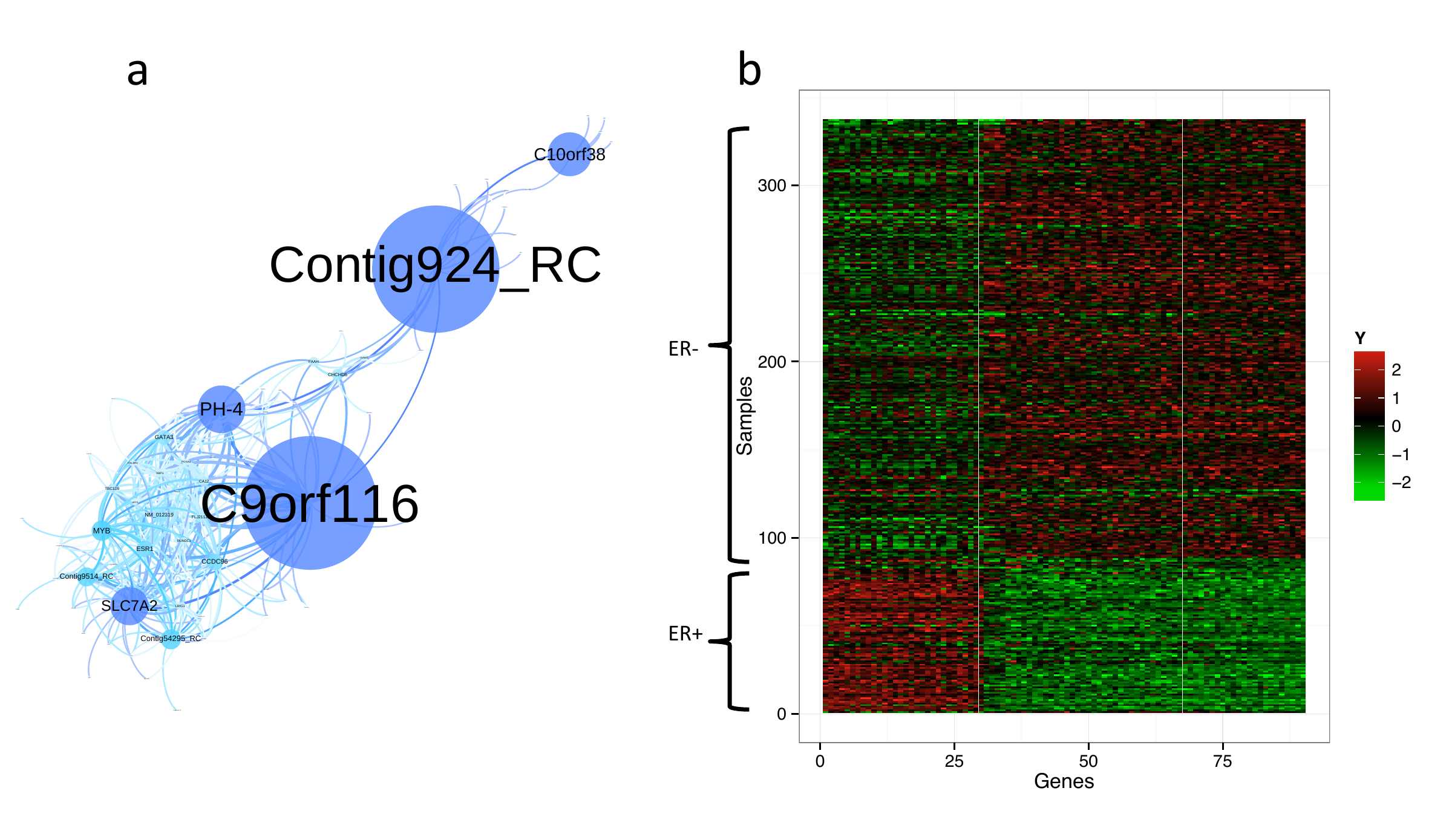}
\end{center}
\vspace{-10mm}
\caption{{\bf Differential ER gene co-expression network and gene expression for ER differential genes.} Panel a: differential ER gene co-expression network, where node size corresponds to betweenness centrality. Panel b: gene expression levels for $90$ genes in the ER differential gene co-expression network.}
\label{fig:Breast-network}
\end{figure}

For the network of differential co-expression across ER+ and ER- samples,  we recovered $90$ genes and $357$ edges that were replicated $>10$ times across the $177$ runs (Figure~\ref{fig:Breast-network}), Supplemental Table S15). We hypothesied that, because the network was differential, the 90 genes may be divided into two sub-groups: a group of genes that are up-regulated in the ER+ samples and down-regulated in the ER- samples, and a group of genes that are down-regulated in the ER+ samples and up-regulated in the ER- samples.
To test this hypothesis, we quantified differential expression levels for the $90$ genes in the original gene expression matrix (Figure~\ref{fig:Breast-network}b).
We found that these genes are indeed stratified into two groups that show distinct expression patterns among the the ER+ and ER- samples. In comparison, the genes in the ER-status specific co-expression networks do not show such a dramatic pattern of differential expression (Supplemental Figures S4).

In the ER differential network, we found that many of the annotated hub genes
play critical roles in breast tumor development. For example, \emph{ESR1}, one of the two genes that encodes the ER protein, is ranked $10$th in this network based on the betweenness centrality criteria across all network genes, suggesting that much of the regulatory activity of tumor development is modulated by interactions with \emph{ESR1}. 
\emph{MYB} encodes the \emph{MYB} proto-oncogene protein, a member of the \emph{MYB} (myeloblastosis) family of transcription factors.
\emph{MYB} is highly expressed in ER+ breast tumors and tumor cell lines~\citep{gonda_myb_estrogen_2008}; a deficit in \emph{MYB} activity leads to death in animals~\citep{ramsay_myb_2008}.  \emph{GATA3} is a transcription factor pivotal in mediating enhancer accessibility at regulatory regions involved in \emph{ESR1}-mediated transcription~\citep{theodorou_gata3_2013} and is particularly useful as a marker for metastatic breast carcinoma~\citep{cimino-gata3_2013}. \emph{LRIG1}, the leucine-rich repeat and immunoglobulin-like domain 1, has three roles: as an \emph{ERBB} negative regulator, as an intestinal stem cell marker, and as a tumor suppressor~\citep{krig_lrig1_2011,wang_lrig1_2013}. \emph{XBP1} is co-expressed with \emph{ESR1}~\citep{andres_ER_XBP_2011} and promotes triple-negative breast cancer by controlling the \emph{HIF1$\alpha$} pathway~\citep{chen_xbp1_2014}. \emph{PH4} has been shown to predictive of response in primary breast cancer patients receiving neoadjuvant chemotherapy~\citep{ph4:2013}.  

\section{Discussion}

In this work, we developed a Bayesian biclustering model based on a
latent factor model. We included a two-component mixture distribution
to allow both sparse and dense representations of the features or
samples to capture structured variation within the gene expression
data. We used the regularized covariance matrix estimated from the
latent factor model to build a Gaussian Markov random field with the
features as nodes in the undirected network. By extracting covariance
matrices corresponding to subsets of components, we were able to
identify gene co-expression networks that were shared across all
samples, unique to a subset of samples, or differential across sample
subsets. 
We applied our methodology to breast tissue gene expression samples
and recovered co-expression networks that are differential across ER+
and ER- tumor types.

Factor analysis methods, including the biclustering approach presented here but extending to many other well-studied models, are statistical tools developed for exploratory analyses of the data. In this work, we have exploited the latent structure in both the factor and the loading matrix to estimate the covariance matrix that is specific to sample subsets. Here we consider tissue type and tumor types, but these methods can be used for any observed binary, categorical, integer, or continuous covariate (e.g., case-control status, batch, sex, age, EBV load). 

Our results show that a number of genes are identified as part of multiple tissue-specific networks. While individual genes may overlap across networks, the interactions of those genes do not. Genes that co-occur in multiple tissue-specific networks are good candidates to test for differential function across tissues. We also will use this approach to study sexual dimorphism, extracting gene networks specific to one sex or differential across the sexes. As in this work, this will greatly improve our power to identify sex-specific trans-eQTLs using (necessarily) unmatched samples.

In this current version of this approach, extracting a covariance matrix specific to a subset of the samples is performed \emph{post hoc}: the linear projection to the latent space is performed in a mostly unsupervised way, although sparsity does add additional structure above SFA-type approaches. As described in the Results, there are multiple categories of networks that we recover. These categories include: gene networks that exist across tissues, gene networks that are unique to specific tissues, gene networks that are specific to specific subsets of samples (e.g., tumor types), and gene networks that exist across multiple tissues but are differentially co-expressed in those tissues. However, this approach is restrictive in that a covariate of interest does not directly inform the projection. Indirectly, we see that the sparsity structure on the samples allows small subsets of the samples to inform projection, but this still relies on a post hoc interpretation of those sample subsets to recover specific network types. Furthermore, it may be the case that, for the covariate we are interested in (e.g., age, sex), we do not see sufficient signal that is uniquely attributable to those samples (e.g., female) to identify a covariance matrix corresponding to the values of interest. We are currently extending this approach so that the linear projection is explicitly informed by the covariate of interest.

\section{Methods}
We extend the following factor analysis model
\begin{align}
    \*{Y=\Lambda X} + \boldsymbol{\epsilon}
\end{align}
where $\*Y \in \Re^{p \times n}$ is the matrix of observed variables, $\*\Lambda \in \Re^{p \times K}$ is the loading matrix,  $\*X \in \Re^{K \times n}$ is the factor matrix, and $\*\epsilon \in \Re^{p \times n}$ is the residual error matrix. We assume $\epsilon_{\cdot,i} \sim \mathcal{N}(0,\*\Psi)$ for $i=1,\dots,n$, where $\*\Psi = \text{diag}(\psi_1,\dots,\psi_p)$.

In previous work~\citep{Gao2012,Gao2013}, a three parameter beta ($\!{TPB}$)~\citep{Armagan2011} prior was used to model the variance of $\*\Lambda$. Here we use $\!{TPB}$ to induce flexible shrinkage to both $\*\Lambda$ and $\*X$. Specifically, we induce three layers of shrinkage---global, factor specific, and local---to both the factors and the loadings. Below, we describe the model structure for $\*\Lambda$; $\*X$ has a prior structure identical to $\*\Lambda$ (for complete details see Appendix A). 

The hierarchical structure for $\*\Lambda$ is written as
\begin{align}
\varrho &\sim \mathcal{TPB}(e,f,\nu),\ \ \ \ 
\zeta_k \sim \mathcal{TPB}(c,d,\frac{1}{\varrho}-1)\notag\\
\varphi_{i,k} &\sim \mathcal{TPB}(a,b,\frac{1}{\zeta_k}-1),\ \ \ \ 
\Lambda_{i,k} \sim \mathcal{N}(0,\frac{1}{\varphi_{i,k}}-1).\label{spec:lam}
\end{align}
Making the substitution $\eta=\frac{1}{\varrho}-1,\ \ \phi_k=\frac{1}{\zeta_k}-1,\ \ \theta_{i,k}=\frac{1}{\varphi_{i,k}}-1$
we get the equivalent hierarchical structure~\citep{Armagan2011}:
\begin{eqnarray}
\gamma &\sim& \Ga(f,\nu),\ \ 
\eta \sim \Ga(e,\gamma),\ \  
\tau_k \sim \Ga(d,\eta), \notag \ \   
\phi_k \sim \Ga(c,\tau_k)\notag\\
\delta_{i,k} &\sim& \Ga(b,\phi_k), \ \   
\theta_{i,k} \sim \Ga(a,\delta_{i,k}) \ \  
\Lambda_{i,k} \sim \mathcal{N}(0,\theta_{i,k}). \label{eq:lambda-element}
\end{eqnarray}

We write the $i^{th}$ row of $\Lambda$ as a multivariate Gaussian distribution, 
\begin{align}
\Lambda_i \sim \!{N}(0,\*\Theta_i)
\end{align}
where $\*\Theta_i=\text{diag}(\theta_{i,1}, \theta_{i,2}, \cdots, \theta_{i,K})$.
We applied a two-component mixture model to jointly model sparse and dense effects by letting $\theta_{i,k}$ be generated from a mixture of sparse and dense components:
\begin{align}
\theta_{i,k}\sim \pi\Ga(a,\delta_{i,k}) + (1-\pi) \delta(\phi_k),
\end{align}
where the hidden variable $Z_k$ indicates whether the loading is sparse (0) or dense (1) and has a beta Bernoulli distribution:
\begin{align}
\pi |\alpha,\beta &\sim Be(\alpha,\beta)\\
Z_k|\pi &\sim \operatorname{Bern}(\pi),\  k=\{1,\dots, K\}.
\end{align}
A variational EM (VEM) algorithm was constructed for fast inference of the parameters (Appendix A); a Markov chain Monte Carlo approach was developed to propose initial states for VEM (Appendix B).




\acks{The authors would like to acknowledge the GTEx Consortium
  members for useful feedback. CG, SZ, and BEE were
  funded by NIH R00 HG006265 and NIH R01 MH101822. CDB was funded by
  NIH R01 MH101822. All code is available at the websites of CG and
  BEE.}


\newpage

\appendix
\section*{Appendix A.}
\label{app:theorem}

\subsection*{The complete biclustering model for BicMix}
We consider the following factor analysis model:
\begin{align}
    \*{Y=\Lambda X} + \boldsymbol{\epsilon},
\end{align}
where $\*Y \in \Re^{p \times n}$ is the matrix of observed variables, $\*\Lambda \in \Re^{p \times K}$ is the loading matrix,  $\*X \in \Re^{K \times n}$ is the factor matrix, and $\*\epsilon \in \Re^{p \times n}$ is the residual error matrix for $p$ genes and $n$ samples. We assume $\epsilon_{\cdot,j} \sim \mathcal{N}(0,\*\Psi)$, where $\*\Psi = \text{diag}(\psi_1,\dots,\psi_p)$.

In previous work~\citep{Gao2012,Gao2013}, a three parameter beta ($\!{TPB}$)~\citep{Armagan2011} prior was used to model the variance of $\*\Lambda$. The three parameter distribution has the form of 
\begin{equation}
f(x:a,b,\phi)=\frac{\Gamma(a+b)}{\Gamma(a)\Gamma(b)}\phi^bx^{b-1}(1-x)^{a-1}\{1+(\phi-1)x\}^{-(a+b)},
\end{equation}
for $x\in (0,1)$, $a>0$, $b>0$ and $\phi>0$. 

Here we use $\!{TPB}$ to induce flexible shrinkage to both $\*\Lambda$ and $\*X$. Specifically, we induce three layers of shrinkage---global, factor specific and local---for both the factors and the loadings. Below, we describe the model structure for $\*\Lambda$ and $\*X$.

\subsection*{Hierarchical structure for $\*\Lambda$}
The hierarchical structure for $\*\Lambda$ is written as
\begin{align}
\varrho &\sim \mathcal{TPB}(e,f,\nu),\\
\zeta_k &\sim \mathcal{TPB}(c,d,\frac{1}{\varrho}-1)\\
\varphi_{i,k} &\sim \mathcal{TPB}(a,b,\frac{1}{\zeta_k}-1),\\
\Lambda_{i,k} &\sim \mathcal{N}(0,\frac{1}{\varphi_{i,k}}-1).\label{spec:lam}
\end{align}
We use the fact that 
\begin{equation}
\label{eq:gamma_beta_relation}
\varphi \sim \mathcal{TPB}(a,b,\nu) \Leftrightarrow \frac{\theta}{\nu} \sim Be^{\prime}(a,b) \Leftrightarrow \theta \sim \mathcal{G}a(a,\delta) \; \mathrm{ and } \; \delta \sim \mathcal{G}a(b,\nu),
\end{equation}
where $Be^{\prime}(a,b)$ and $\mathcal{G}a$ indicate an inverse beta and a gamma distribution, respectively.
Making the substitution  $\eta=\frac{1}{\varrho}-1,\ \ \phi_k=\frac{1}{\zeta_k}-1,\ \ \theta_{i,k}=\frac{1}{\varphi_{i,k}}-1$,
we get the equivalent hierarchical structure~\citep{Armagan2011}:
\begin{eqnarray}
\gamma &\sim& \Ga(f,\nu),\\
\eta &\sim& \Ga(e,\gamma),\\  
\tau_k &\sim& \Ga(d,\eta),\\   
\phi_k &\sim& \Ga(c,\tau_k),\\
\delta_{i,k} &\sim& \Ga(b,\phi_k),\\   
\theta_{i,k}&\sim& \Ga(a,\delta_{i,k}), \\  
\Lambda_{i,k} &\sim& \mathcal{N}(0,\theta_{i,k}). \label{eq:lambda-element}
\end{eqnarray}

We applied a two-component mixture model to jointly model possibly dense confounding effects by letting $\theta_{i,k}$ be generated from a mixture of sparse and dense components:
\begin{align}
\theta_{i,k}\sim \pi\Ga(a,\delta_{i,k}) + (1-\pi) \delta(\phi_k),
\end{align}
where the hidden variable $Z_k$, which indicates whether or not the loading is sparse (1) or dense (0), is generated from the following beta Bernoulli distribution:
\begin{align}
\pi |\alpha,\beta &\sim Be(\alpha,\beta)\\
Z_k|\pi &\sim \operatorname{Bern}(\pi),\  k=\{1,\dots, K\}.
\end{align}

\subsection*{Hierarchical structure for $\*X$}
Similarly, the hierarchical structure for $\*X$ is:
\begin{eqnarray}
\varphi &\sim& \Ga(f_X,\xi), \\
\chi &\sim& \Ga(e_X,\varphi),\\  
\kappa_k &\sim& \Ga(d_X,\chi),\\ 
\omega_k &\sim& \Ga(c_X,\kappa_k)\\
\rho_{k,j} &\sim& \Ga(b_X,\omega_k),\\ 
\sigma_{k,j} &\sim& \pi\Ga(a_X,\rho_{k,j}) + (1-\pi) \delta(\omega_k)\\
x_{k,j} &\sim& \mathcal{N}(0,\sigma_{i,k})
\end{eqnarray} 
with $\sigma_{k,j}$ being generated from a two component mixture.
Here, the hidden variable $O_k$, which indicates whether or not the factor is sparse (1) or dense (0), is generated from the following beta Bernoulli distribution:
\begin{align}
\pi_X |\alpha_X,\beta_X &\sim Be(\alpha_X,\beta_X)\\
O_k|\pi_X &\sim \operatorname{Bern}(\pi_X),\  k=\{1,\dots, K\}.
\end{align}

\subsection*{Variational expectation maximization}

Extending previous work~\citep{Gao2013}, the posterior probability $\!P=p(\*{\Lambda,X,Z,O,\Theta|Y})$ is written as:
\begin{align}
\!P &\propto p(\*{Y|\Lambda,X})p(\*{\Lambda|Z,\Theta_\Lambda})p(\*{X|O,\Theta_X})p(\*{Z|\Theta_\Lambda})p(\*{O|\Theta_X})p(\*\Theta_\Lambda)p(\*\Theta_X)\notag\\
&\propto p(\*{Y|\Lambda,X}) \!{P}(\*\Lambda) \!P(\*X),
\end{align}
where we have used $\*\Theta_\Lambda$ and $\*\Theta_X$ to denote the set of parameters related to $\*\Lambda$ and $\*X$ respectively. Then,
\begin{align}
\!P(\*\Lambda) &= p(\*{\Lambda|Z,\Theta_\Lambda})p(\*{Z|\Theta_\Lambda})p(\*\Theta_\Lambda)\\
&= \[\prod_{i=1}^p\prod_{k=1}^K
\N(\Lambda_{i,k} | \theta_{i,k})\Ga(\theta_{i,k} | a,\delta_{i,k})\Ga(\delta_{i,k}|b,\phi_k)
\]^{\mathds{1}_{Z_k=1}}\notag\\
&\times \[\prod_{i=1}^p\prod_{k=1}^K \N(\Lambda_{i,k} | \phi_k) \]^{\mathds{1}_{Z_k=0}}
\[ \prod_{k=1}^K \mathcal{B}ern(Z_k|\pi)\] \mathcal{B}eta(\pi|\alpha,\beta)\notag
\notag\\
&\times
\[\prod_{k=1}^K \Ga(\phi_k|c,\tau_k)\Ga(\tau_k | d,\eta)\]\Ga(\eta | e,\gamma)\Ga(\gamma|f,\nu)\notag
\end{align}
and
\begin{align}
\!P(\*X) &= p(\*{X|O,\Theta_X})p(\*{O|\Theta_X})p(\*\Theta_X)\\
&= \[\prod_{k=1}^K\prod_{j=1}^n
\N(x_{k,j} | \sigma_{k,j})\Ga(\sigma_{k,j} | a_X,\rho_{k,j})\Ga(\rho_{k,j}|b_X,\omega_k)
\]^{\mathds{1}_{O_k=1}}\notag\\
&\times \[\prod_{k=1}^K\prod_{j=1}^n \N(x_{k,j} | \omega_k) \]^{\mathds{1}_{O_k=0}}
\[ \prod_{k=1}^K \mathcal{B}ern(o_k|\pi_X)\] \mathcal{B}eta(\pi_X|\alpha_X,\beta_X)\notag
\notag\\
&\times
\[\prod_{k=1}^K \Ga(\omega_k|c_X,\kappa_k)\Ga(\kappa_k | d_X,\chi)\]\Ga(\chi | e_X,\varphi)\Ga(\varphi|f_X,\xi)\notag
\end{align}

\subsection*{Parameters specific to $\*\Lambda$}
To derive the variational EM algorithm, we expand the posterior probability (Equation~(18)) and write the expected complete log likelihood for parameters related to $\*\Lambda$,  $Q(\*\Theta_\Lambda)=\^{\ell_c(\*\Theta_\Lambda, \*\Lambda|\*Z,\*X,\*Y)}$ as:
\begin{align}
Q(\*\Theta_\Lambda) &\propto 
\sum_{i=1}^p\sum_{j=1}^n\^{\log p(y_{i,j}|\*\Lambda,\*X,\*\Theta_\Lambda,\*Z)}\\
&+\sum_{i=1}^p\sum_{k=1}^K\^{p(Z_k|\*\Theta_\Lambda)\log p(\Lambda_{i,k}|\*\Theta_\Lambda,Z_k)}+\log p(\*\Theta_\Lambda)\notag\\
&\propto
-\frac{p}{2}\ln|\*\Psi|-\sum_{i=1}^p\sum_{j=1}^n\frac{\(y_{i,j}-\sum_{k=1}^K\Lambda_{i,k}\^{x_{k,j}}\)^2}{2\psi_{i,i}}
+\sum_{i=1}^p\sum_{k=1}^K
\^{1-z_k}\left\{
-\frac{1}{2}\ln\phi_k-\frac{\Lambda_{i,k}^2}{2\phi_k}
\right\}\notag\\
&+\sum_{i=1}^p\sum_{k=1}^K
\^{z_k}\left\{
-\frac{1}{2}\ln\theta_{i,k}-\frac{\Lambda_{i,k}^2}{2\theta_{i,k}}
+a\ln\delta_{i,k}+(a-1)\ln\theta_{i,k}-\delta_{i,k}\theta_{i,k}
\right\}\notag\\
&+\sum_{i=1}^p\sum_{k=1}^K\^{z_k}\left\{b\ln\phi_k+(b-1)\ln\delta_{i,k}-\phi_k\delta_{i,k}\right\}
+\sum_{k=1}^K\left\{\^{z_k}\ln\pi+(1-\^{z_k})\ln(1-\pi)\right\}\notag\\
&+\sum_{k=1}^K\left\{
c\ln\tau_k+(c-1)\ln\phi_k-\tau_k\phi_k+d\ln\eta+(d-1)\ln\tau_k-\eta\tau_k
\right\}\notag\\
&+e\ln\gamma+(e-1)\ln\eta-\gamma\eta+f\ln\nu+(f-1)\ln\gamma-\nu\gamma +\alpha \ln \pi + \beta \ln (1-\pi)\notag
\end{align}
where we have used $\^{X}$ to represent the expected value of $X$.

We obtain the MAP estimates $\hat{\*\Theta_\Lambda}=\argmax_{\*\Theta_\Lambda} Q(\*\Theta_\Lambda)$. Specifically, we solve equation $\frac{\partial Q(\*\Theta_\Lambda)}{\partial\*\Theta_\Lambda} = 0$ to find the closed form of their MAP estimates.
The MAP estimate for the $i$th row of $\*\Lambda$, $\Lambda_{i,\cdot}$, in matrix form is:
\begin{align}
\hat{\Lambda}_{i,\cdot} = Y_{i,\cdot}\Psi_{i,i}^{-1}\*X^T\(\^{\*X\Psi_{i,i}^{-1}\*X^T}+\*{\^{\mathds{Z}}}\*\Theta_i^{-1}+(\*I-\^{\*{\mathds{Z}}})\*\Phi^{-1}\)^{-1}\label{eq:VEM-lambda}
\end{align}
where 
\begin{align}{\label{eq:thetai}}
\*\Theta_i =
 \begin{pmatrix}
  \theta_{i,1} & 0 & \cdots & 0 \\
  0 & \theta_{i,2} & \cdots & 0 \\
  \vdots  & \vdots  & \ddots & \vdots  \\
  0 & 0 & \cdots & \theta_{i,k}
\end{pmatrix},\ \ 
\*\Phi =
 \begin{pmatrix}
  \phi_{1} & 0 & \cdots & 0 \\
  0 & \phi_{2} & \cdots & 0 \\
  \vdots  & \vdots  & \ddots & \vdots  \\
  0 & 0 & \cdots & \phi_{k}
\end{pmatrix},\ \ 
\end{align}
and
\begin{align}
\*{\mathds{\^Z}} =
 \begin{pmatrix}
  \^{z_{1}} & 0 & \cdots & 0 \\
  0 & \^{z_{2}} & \cdots & 0 \\
  \vdots  & \vdots  & \ddots & \vdots  \\
  0 & 0 & \cdots & \^{z_{k}}
\end{pmatrix}
\end{align}
and $\*I$ is the identity matrix.

Parameter $\theta_{i,k}$ has a generalized inverse Gaussian conditional probability ~\citep{Gao2013,Armagan2011}, and its MAP estimate is 
\begin{eqnarray}
\hat{\theta}_{i,k}&=&\frac{2a-3+\sqrt{(2a-3)^2+8\Lambda_{i,k}^2\delta_{i,k}}}{4\delta_{i,k}}\label{eq:VEM-theta}
\end{eqnarray}
The MAP estimate for $\delta_{i,k}$ is:
\begin{eqnarray}
\hat{\delta}_{i,k}&=&\frac{a+b-1}{\theta_{i,k}+\phi_k}\label{eq:VEM-delta}
\end{eqnarray}
Parameter $\phi_k$ generates both the sparse and dense components, and its MAP estimate takes the form of
\begin{align}
\hat{\phi}_k&=\frac{H+\sqrt{H^2+M T}}{M}\label{eq:VEM-phi}
\end{align}
where
\begin{align}
H &= pb\z_k+c-1-\frac{p}{2}(1-\z_k)\\
M &= 2\left(\z_k\sum_{i=1}^p\delta_{i,k}+\tau_k\right)\\
T &= \sum_{i=1}^p\Lambda_{i,k}^2.
\end{align}
The following parameters have similar updates to $\delta_{i,k}$, which have natural forms because of conjugacy:
\begin{eqnarray}
\hat{\tau}_k&=&\frac{c+d-1}{\phi_k+\eta}\label{eq:VEM-tau}\\
\hat{\eta}&=&\frac{Kd+e-1}{\gamma+\sum_k\tau_k}\label{eq:VEM-eta}\\
\hat{\gamma}&=&\frac{e+f-1}{\eta+\nu}\label{eq:VEM-gamma}
\end{eqnarray}

The expected value of $Z_{k}|\*\Theta$ is computed as:
\begin{align}
\langle z_{k}|\*\Theta_\Lambda\rangle&= p(Z_k=1|\*\Theta_\Lambda)\label{eq:VEM-z}\\\
&=\frac{\pi\prod_{i=1}^p\N(\Lambda_{i,k} | \theta_{i,k})\Ga(\theta_{i,k} | a,\delta_{i,k})\Ga(\delta_{i,k}|b,\phi_k)}{(1-\pi)(\prod_{i=1}^p \N(\Lambda_{i,k} | \phi_k))+\pi\prod_{i=1}^p\N(\Lambda_{i,k} | \theta_{i,k})\Ga(\theta_{i,k} | a,\delta_{i,k})\Ga(\delta_{i,k}|b,\phi_k)}notag
\end{align}

The prior on the indicator variable for sparse and dense components, $\pi$, has a beta distribution, and its geometric mean is the following:
\begin{align}
\langle \ln \pi\rangle&=\psi\left(\sum_{k=1}^K{\z_k}+\alpha\right)-\psi\left(K+\alpha+\beta\right) \label{eq:VEM-pi}
\end{align}
where $\psi$ is the digamma function.
\subsection*{Parameters specific to $\*X$}
Similarly, the expected complete log likelihood for parameters related to $\*X$ takes the following form:

\begin{align}
Q(\*\Theta_X) &\propto 
\sum_{i=1}^p\sum_{j=1}^n\^{\log p(y_{i,j}|\*\Lambda,\*X,\*\Theta_X,\*O)}\\
&+\sum_{k=1}^K\sum_{j=1}^n\^{p(O_k|\*\Theta_X)\log p(X_{k,j}|\*\Theta_X,O_k)}+\log p(\*\Theta_X)\notag\\
&\propto
-\frac{p}{2}\ln|\*\Psi|-\sum_{i=1}^p\sum_{j=1}^n\frac{\(y_{i,j}-\sum_{k=1}^K\Lambda_{i,k}\^{x_{k,j}}\)^2}{2\psi_{i,i}}
+\sum_{k=1}^K\sum_{j=1}^n
\^{1-o_k}\left\{
-\frac{1}{2}\ln\omega_k-\frac{\^{x_{k,j}^2}}{2\omega_k}
\right\}\notag\\
&+\sum_{k=1}^K\sum_{j=1}^n
\^{o_k}\left\{
-\frac{1}{2}\ln\sigma_{k,j}-\frac{\^{x_{k,j}^2}}{2\sigma_{k,j}}
+a_X\ln\rho_{k,j}+(a_X-1)\ln\sigma_{k,j}-\rho_{k,j}\sigma_{k,j}
\right\}\notag\\
&+\sum_{k=1}^K\sum_{j=1}^n\^{o_k}\left\{b_X\ln\omega_k+(b_X-1)\ln\rho_{k,j}-\omega_k\rho_{k,j}\right\}
+\sum_{k=1}^K\left\{\^{o_k}\ln\pi_X+(1-\^{o_k})\ln(1-\pi_X)\right\}\notag\\
&+\sum_{k=1}^K\left\{
c_X\ln\kappa_k+(c_X-1)\ln\omega_k-\kappa_k\omega_k+d_X\ln\chi+(d_X-1)\ln\kappa_k-\chi\kappa_k
\right\}\notag\\
&+e_X\ln\varphi+(e_X-1)\ln\chi-\varphi\chi+f_X\ln\xi+(f_X-1)\ln\varphi-\xi\varphi +\alpha_X \ln \pi_X + \beta_X \ln (1-\pi_X)\notag
\end{align}
To simplify the calculation, we assumed that the joint distribution $p(o_k,x_{k,j})$ factorizes to $p(o_k)p(x_{k,j})$. 

The closed form solutions for the parameters related to $\*X$ are:
\begin{align}
\^{\*X_{\cdot,j}} &= (\*\Lambda^T\*\Psi^{-1}\*\Lambda+\^{\mathds{O}}\*\Sigma_j^{-1}+\(\*I-\^{\mathds{O}}\)\*\Omega^{-1})^{-1}\*\Lambda^T\*\Psi^{-1}Y_{\cdot,j},\label{eq:VEM-x}
\end{align}
where 
\begin{align}\label{eq:sigmaj}
\*\Sigma_j =
 \begin{pmatrix}
  \sigma_{1,j} & 0 & \cdots & 0 \\
  0 & \sigma_{2,j} & \cdots & 0 \\
  \vdots  & \vdots  & \ddots & \vdots  \\
  0 & 0 & \cdots & \sigma_{k,j}
\end{pmatrix},\ \ 
\*\Omega =
 \begin{pmatrix}
  \omega_{1} & 0 & \cdots & 0 \\
  0 & \omega_{2} & \cdots & 0 \\
  \vdots  & \vdots  & \ddots & \vdots  \\
  0 & 0 & \cdots & \omega_{k}
\end{pmatrix},\ \ 
\end{align}
and
\begin{align}
\*{\mathds{O}} =
 \begin{pmatrix}
  \^{o_{1}} & 0 & \cdots & 0 \\
  0 & \^{o_{2}} & \cdots & 0 \\
  \vdots  & \vdots  & \ddots & \vdots  \\
  0 & 0 & \cdots & \^{o_{k}}
\end{pmatrix}
\end{align}

In these equations,
\begin{eqnarray}
\hat{\sigma}_{k,j}&=&\frac{2a_X-3+\sqrt{(2a_X-3)^2+8\^{x_{k,j}^2}\rho_{k,j}}}{4\rho_{k,j}},\label{eq:VEM-sigma}
\end{eqnarray}

\begin{eqnarray}
\hat{\rho}_{k,j}&=&\frac{a_X+b_X-1}{\sigma_{k,j}+\omega_k},\label{eq:VEM-rho}
\end{eqnarray}

\begin{align}
\hat{\omega}_k&=\frac{H+\sqrt{H^2+M T}}{M},\label{eq:VEM-omega}
\end{align}
where
\begin{align}
H &= nb_X\o_k+c_X-1-\frac{n}{2}(1-\o_k)\\
M &= 2\left(\o_k\sum_{j=1}^n\rho_{k,j}+\kappa_k\right)\\
T &= \sum_{j=1}^n\^{x_{k,j}^2}.
\end{align}

We also need to compute the expected value of $\*X\Psi_{i,i}^{-1}\*X^T$:
\begin{align}\label{eq:EXX}
\^{\*X\Psi_{i,i}^{-1}\*X^T} = \Psi_{i,i}^{-1}\(\^{\*X}\^{\*X}^T+\*\Sigma_X\)
\end{align}
where $\*\Sigma_X$ denotes the covariance matrix of $\*X$.

Finally:
\begin{eqnarray}
\hat{\kappa}_k&=&\frac{c_X+d_X-1}{\omega_k+\chi}\label{eq:VEM-kappa}\\
\hat{\chi}&=&\frac{Kd_X+e_X-1}{\varphi+\sum_k\kappa_k}\label{eq:VEM-chi}\\
\hat{\varphi}&=&\frac{e_X+f_X-1}{\chi+\xi}.\label{eq:VEM-varphi}
\end{eqnarray}

Now we consider the parameters $O_k$:
\begin{align}
\langle O_{k}|\*\Theta_X\rangle&= p(O_k=1|\*\Theta_X) \label{eq:VEM-o}\\
&=\frac{\pi\prod_{j=1}^n\N(X_{k,j} | \sigma_{k,j})\Ga(\sigma_{k,j} | a_X,\rho_{k,j})\Ga(\rho_{k,j}|b_X,\omega_k)}{(1-\pi)(\prod_{j=1}^n \N(X_{k,j} | \omega_k))+\pi\prod_{j=1}^n\N(X_{k,j} | \sigma_{k,j})\Ga(\sigma_{k,j} | a_X,\rho_{k,j})\Ga(\rho_{k,j}|b_X,\omega_k)}.\notag
\end{align}

\begin{align}
\langle \ln \pi_X\rangle&=\psi\left(\sum_{k=1}^K{\o_k}+\alpha_X\right)-\psi\left(K+\alpha_X+\beta_X\right) \label{eq:VEM-varpi}
\end{align}
and finally, assuming that the residual precision has a conjugate
prior, $\frac{1}{\Psi_{i,i}} \sim \!Ga(1,1)$, then we have
\begin{align}
\*\Psi = \text{diag}\(\frac{\*{YY^T-2Y\^{X^T}\Lambda^T+\Lambda\^{XX^T}\Lambda^T}+2\*I}{n+2}\). \label{eq:VEM-psi}
\end{align}

\subsection*{VEM algorithm}\label{sec:VEM}
To summarize the description above, we write the complete VEM
algorithm for parameter updates:

\begin{algorithm}[H]
\DontPrintSemicolon
\KwData{$p\times n$ Gene expression matrix, $K$, $n\_itr$}
\KwResult{$p\times K$ and $K\times n$ matrices with mixture of sparse and dense components}
\Begin{
\textbf{Initialize starting values:}\;
$a,b,c,d,e,f,a_X,b_X,c_X,d_X,e_X,f_X \leftarrow 0.5$\;
$\alpha,\beta,\alpha_X,\beta_X \leftarrow 1$\;
Sample $\eta,\gamma,\chi,\varphi \leftarrow$ $\!Ga(1,1)$\;
Sample $\pi \leftarrow$ $\!Beta(\alpha,\beta)$, $\pi_X \leftarrow$ $\!Beta(\alpha_X,\beta_X)$,\;
\For{$i\leftarrow 1$ \KwTo $p$}{
Sample $\psi_{i,i} \leftarrow$ $\!Ga(1,1)$
}
\For{$k\leftarrow 1$ \KwTo $K$}{
  Sample $z_k \leftarrow$ $\!Bern(\pi)$, $o_k \leftarrow$ $\!Bern(\pi_X)$\;
  Sample $\phi_k,\tau_k,\omega_k,\kappa_k \leftarrow$ $\!Ga(1,1)$\;
\For{$i\leftarrow 1$ \KwTo $p$}{
  Sample $\Lambda_{i,k} \leftarrow$ $\!N(0,1)$,
  Sample $\theta_{i,k},\delta_{i,k} \leftarrow$ $\!Ga(1,1)$\;
}
\For{$j\leftarrow 1$ \KwTo $n$}{
  Sample $X_{k,j} \leftarrow$ $\!N(0,1)$,
  Sample $\sigma_{k,j},\rho_{k,j} \leftarrow$ $\!N(0,1)$
}
}
\textbf{Begin iterations:}\;
\For{$t\leftarrow 1$ \KwTo $n\_itr$}{
  \textbf{Update parameters that are specific to $\*\Lambda$:}\;
    \For{$i\leftarrow 1$ \KwTo $p$}{
      Update $\Lambda_{i,\cdot} \leftarrow$ equation (\ref{eq:VEM-lambda})\;
      \For{$k\leftarrow 1$ \KwTo $K$}{
      Update $\theta_{i,k} \leftarrow$ equation (\ref{eq:VEM-theta}),
      $\delta_{i,k} \leftarrow$ equation (\ref{eq:VEM-delta})
    }
  }
  \For{$k\leftarrow 1$ \KwTo $K$}{
    Update $\phi_k \leftarrow$ equation (\ref{eq:VEM-phi}),
    $\tau_k \leftarrow$ equation (\ref{eq:VEM-tau}),
    $z_k \leftarrow$ equation (\ref{eq:VEM-z})
  }
  Update $\eta \leftarrow$ equation (\ref{eq:VEM-eta}),
  $\gamma \leftarrow$ equation (\ref{eq:VEM-gamma}),
  $\pi \leftarrow$ equation (\ref{eq:VEM-pi})\;
  \textbf{Update parameters that are specific to $\*X$:}\;

  \For{$j\leftarrow 1$ \KwTo $n$}{
    Update $X_{\cdot,j} \leftarrow$ equation (\ref{eq:VEM-x})\;
    \For{$k\leftarrow 1$ \KwTo $K$}{
      Update $\sigma_{k,j} \leftarrow$ equation (\ref{eq:VEM-sigma}), 
      $\rho_{k,j} \leftarrow$ equation (\ref{eq:VEM-rho})\;
    }
  }
  \For{$k\leftarrow 1$ \KwTo $K$}{
    Update $\omega_k \leftarrow$ equation (\ref{eq:VEM-omega}),
    $\kappa_k \leftarrow$ equation (\ref{eq:VEM-kappa}),
    $o_k \leftarrow$ equation (\ref{eq:VEM-o})
  }
  Update $\chi \leftarrow$ equation (\ref{eq:VEM-chi}),
  $\varphi \leftarrow$ equation (\ref{eq:VEM-varphi}),
  $\pi_X \leftarrow$ equation (\ref{eq:VEM-varpi})\;
  \For{$i\leftarrow 1$ \KwTo $p$}{
    Update $\psi_{i,i} \leftarrow$ equation (\ref{eq:VEM-psi})\;
  }
}
Output $\*\Lambda$, $\*X$, $Z$, $O$\;
}
\caption{Variational expectation maximization for BicMix\label{VEM}}
\end{algorithm}

\section*{Appendix B.}

\subsection*{MCMC: conditional distributions of BicMix parameters}

We derive below the conditional distributions that capture the MCMC approach that we implemented for BicMix. In our manuscript, we used MCMC to compute the warm start parameter settings in the simulations.

\subsection*{Conditional distributions for parameters related to $\*\Lambda$}
We updated the loading matrix $\*\Lambda$ one row at a time, where
each row consists of values across the $K$ components. The $i$th row of
the loading matrix, $\Lambda_{i,\cdot}$, has the following posterior
distribution,
\begin{align}
\Lambda_{i,\cdot}|Y_{i,\cdot}\*X,\*\Theta_i,\Psi_{i,i} &\sim \!N\(Y_{i,\cdot}\Psi_{i,i}^{-1}\*X^T\({\*X\Psi_{i,i}^{-1}\*X^T}+\*V_i^{-1}\)^{-1},\*X\Psi_{i,i}^{-1}\*X^T+\*V_i^{-1}\)\label{eq:post-lambda}
\end{align}
where $\*V_i$ is a $K \times K$ diagonal matrix. If we use $V_{i,k,k}$
to denote the $(k,k)$th element for $V_i$, then we sample
$V_{i,k,k}$ and its related parameters as follows:
\begin{align}
 V_{i,k,k} = \left\{ \begin{array}{ll}\label{eq:post-vkk}
\theta_{i,k} & \mbox{if $Z_k = 1$};\\
\phi_k & \mbox{if $Z_k = 0$}.\end{array} \right.
\end{align}
We sample values for the parameters conditional on sparse and dense state as follows. If $Z_k=1$,
\begin{align}
&\theta_{i,k}|\Lambda_{i,k},\delta_{i,k} \sim \mathcal{GIG}\left(a-\frac{1}{2},2\delta_{i,k},\Lambda_{i,k}^2\right) \label{eq:post-theta}\\
&\delta_{i,k}|\theta_{i,k},\phi_{k} \sim \mathcal{G}a(a+b,\theta_{i,k}+\phi_k) \label{eq:post-delta}\\
&\phi_k|\delta_{i,k},\tau_{k} \sim \mathcal{G}a\(pb+c,\sum_{i=1}^p\delta_{i,k}+\tau_k\).\label{eq:post-sparse-phi}
\end{align}

If $Z_k = 0$,
\begin{align}
&\phi_k|\tau_{k},\Lambda_{i,k} \sim  \mathcal{GIG}\left(c-\frac{p}{2},2\tau_k,\sum_{i=1}^p \Lambda_{i,k}^2\right). \label{eq:post-dense-phi}
\end{align}

The following parameters are not sparse or dense component specific; they each have a gamma conditional distribution because of conjugacy:
\begin{align}
\tau_k|\phi_k,\eta &\sim \mathcal{G}a\(c+d,\phi_k+\eta\) \label{eq:post-tau}\\
\eta|\gamma,\tau_k &\sim \mathcal{G}a\(Kd+e,\gamma+\sum_{k=1}^K\tau_k\) \label{eq:post-eta}\\
\gamma|\eta,\nu &\sim \mathcal{G}a(e+f,\eta+\nu). \label{eq:post-gamma}
\end{align}

The conditional probability for $Z_k$ has a Bernoulli distribution:
\begin{align}
p(Z_k=1|\*\Theta_\Lambda) = \frac{\pi\prod_{i=1}^p\N(\Lambda_{i,k} | \theta_{i,k})\Ga(\theta_{i,k} | a,\delta_{i,k})\Ga(\delta_{i,k}|b,\phi_k)}{(1-\pi)(\prod_{i=1}^p \N(\Lambda_{i,k} | \phi_k))+\pi\prod_{i=1}^p\N(\Lambda_{i,k} | \theta_{i,k})\Ga(\theta_{i,k} | a,\delta_{i,k})\Ga(\delta_{i,k}|b,\phi_k)}.\notag
\end{align}

Let $p_Z = p\(Z_k=1|\*\Theta_\Lambda\)$; then the conditional probability for $Z_k$ is 
\begin{align}
Z_k|p_Z &\sim \!{B}ern(p_Z). \label{eq:post-z}
\end{align}

The mixing proportion $\pi$ has a beta conditional probability: 
\begin{align}
\pi|\alpha,\beta,Z_k \sim \mathcal{B}eta\(\alpha+\sum_{k=1}^K \mathds{1}_{Z_k=1},K-\sum_{k=1}^K\mathds{1}_{Z_k=0}+\beta\) \label{eq:post-pi}
\end{align}
where $\mathds{1}$ is the indicator function.

\subsection*{Conditional distributions for parameters related to $\*X$}
We updated the factor matrix $\*X$ one column at a time, where each
column consists of values across the $K$ components; the $j$th column
of the factor matrix, $X_{\cdot,j}$, has the following posterior
distribution,
\begin{align}
X_{\cdot,j}|Y_{\cdot,j},\*\Lambda,\*\Sigma_j,\*\Psi &\sim \!N\((\*\Lambda^T\*\Psi^{-1}\*\Lambda+\*W_j^{-1})^{-1}\*\Lambda^T\*\Psi^{-1}Y_{\cdot,j},\*\Lambda^T\*\Psi^{-1}\*\Lambda+\*W_j^{-1}\),\label{eq:post-x}
\end{align}
where $\*W_j$ is a $K \times K$ diagonal matrix. If we use $W_{j,k,k}$
to denote the $(k,k)$th element for $W_j$, then we sample the value of
$W_{j,k,k}$ as follows:
\begin{align}
 W_{j,k,k} = \left\{ \begin{array}{ll}\label{eq:post-wkk}
\sigma_{k,j} & \mbox{if $O_k = 1$};\\
\omega_k & \mbox{if $O_k = 0$}.\end{array} \right.
\end{align}

We sample the values of the parameters conditional on the sparse and dense state as follows. If $O_k = 1$
\begin{align}
&\sigma_{k,j}|X_{k,j},\rho_{k,j} \sim \mathcal{GIG}\left(a_X-\frac{1}{2},2\rho_{k,j},X_{k,j}^2\right) \label{eq:post-sigma}\\
&\rho_{k,j}|\Sigma_{k,j},\omega_{k} \sim \mathcal{G}a(a_X+b_X,\rho_{k,j}+\omega_k) \label{eq:post-rho}\\
&\omega_k|\rho_{k,j},\omega_{k} \sim \mathcal{G}a\(nb_X+c_X,\sum_{j=1}^n\rho_{k,j}+\omega_k\). \label{eq:post-sparse-omega}
\end{align}
If $O_k = 0$
\begin{align}
&\omega_k|\kappa_{k},X_{k,j} \sim  \mathcal{GIG}\left(c_X-\frac{n}{2},2\kappa_k,\sum_{j=1}^n x_{k,j}^2\right). \label{eq:post-dense-omega}
\end{align}
The following parameters are not sparse or dense component specific;
they each have a gamma conditional distribution because of conjugacy:
\begin{align}
&\kappa_k \sim \!Ga\(c_X+d_X,\omega_k+\chi\) \label{eq:post-kappa}\\
&\chi \sim \!{G}a\(Kd_X+e_X,\varphi+\sum_k\kappa_k\) \label{eq:post-chi}\\
&\varphi \sim \!Ga\(e_X+f_X,\chi+\xi\). \label{eq:post-varphi}
\end{align}

The conditional probability for $O_k$ has a Bernoulli distribution:
\begin{align}
p(O_k=1|\*\Theta_X) = \frac{\pi\prod_{j=1}^n\N(X_{k,j} | \sigma_{k,j})\Ga(\sigma_{k,j} | a_X,\rho_{k,j})\Ga(\rho_{k,j}|b_X,\omega_k)}{(1-\pi)(\prod_{j=1}^n \N(X_{k,j} | \omega_k))+\pi\prod_{j=1}^n\N(X_{k,j} | \sigma_{k,j})\Ga(\sigma_{k,j} | a_X,\rho_{k,j})\Ga(\rho_{k,j}|b_X,\omega_k)}.
\end{align}

Let $p_X = p(O_k=1|\*\Theta_X)$; then the conditional probability for $O_k$ is 
\begin{align}
O_k|p_X &\sim \!{B}ern(p_X). \label{eq:post-o}
\end{align}

The mixing proportion $\pi$ has a beta conditional probability: 
\begin{align}
\pi_X|\alpha_X,\beta_X,O_k \sim \mathcal{B}eta\(\alpha_X+\sum_{k=1}^K \mathds{1}_{O_k=1},K-\sum_{k=1}^K\mathds{1}_{O_k=0}+\beta_X\) \label{eq:post-varpi}
\end{align}
where $\mathds{1}$ is the indicator function.
Finally, we have,
\begin{align}
\psi_{i,i} \sim \!{IG}\(\frac{n}{2}+1,\frac{\sum_{j=1}^n \left(y_{i,j}-\sum_{k=1}^K \Lambda_{i,k}x_{k,j}\right)^2}{2}+1\). \label{eq:post-psi}
\end{align}

\subsection*{MCMC for parameter estimation in BicMix}
We implemented the following MCMC algorithm for sampling the parameters of the BicMix model.

\begin{algorithm}[H]
\DontPrintSemicolon
\KwData{$p\times n$ Gene expression matrix, $K$, $n\_itr$}
\KwResult{$p\times K$ and $K\times n$ matrices with both sparse and dense components}
\Begin{
\textbf{Initialization:}\;
Sample the initial values of the parameters as desribed
in Section (\ref{sec:VEM})\;
\textbf{Begin iterations:}\;
\For{$t\leftarrow 1$ \KwTo $n\_itr$}{
  \textbf{Sample parameters specific to $\*\Lambda$:}\;
  \For{$i\leftarrow 1$ \KwTo $p$}{
    \For{$k\leftarrow 1$ \KwTo $K$}{
      Sample $V_{i,k,k}$ according to equation (\ref{eq:post-vkk}) 
    }
    Sample $\Lambda_{i,\cdot}$ according to equation (\ref{eq:post-lambda})\;
  }
  \For{$k\leftarrow 1$ \KwTo $K$}{
    \If { $z_k = 1 $}{
      Sample $\phi_k$ according to equation
      (\ref{eq:post-sparse-phi})\;
      \For{$i\leftarrow 1$ \KwTo $p$}{
        Sample $\theta_{i,k}$ according to equation  (\ref{eq:post-theta}),
      $\delta_{i,k}$ according to equation (\ref{eq:post-delta}),
    }
  }
  \If { $z_k = 0 $}{
    Sample $\phi_k$ according to equation (\ref{eq:post-dense-phi})\;
  }
}  
 \For{$k\leftarrow 1$ \KwTo $K$}{
   Sample $\tau_k$ according to equation (\ref{eq:post-tau}),
   $z_k$ with equation (\ref{eq:post-z}) 
  }
  Sample $\eta$ according to equation (\ref{eq:post-eta}),
  $\gamma$ with equation (\ref{eq:post-gamma}),
  $\pi$ with equation (\ref{eq:post-pi})\;

  \textbf{Sample parameters specific to $\*X$:}\;
  \For{$j\leftarrow 1$ \KwTo $n$}{
    \For{$k\leftarrow 1$ \KwTo $K$}{
      Sample $W_{j,k,k}$ according to equation (\ref{eq:post-wkk}) 
    }
    Sample $X_{\cdot,j}$ according to equation (\ref{eq:post-x})\;
  }
  \For{$k\leftarrow 1$ \KwTo $K$}{
    \If { $o_k = 1 $}{
      Sample $\omega_k$ according to equation
      (\ref{eq:post-sparse-omega})\;
      \For{$j\leftarrow 1$ \KwTo $n$}{
        $\sigma_{k,j}$ according to equation  (\ref{eq:post-sigma}),
      $\rho_{k,j}$ according to equation (\ref{eq:post-rho}),
    }
  }
  \If { $o_k = 0 $}{
    Sample $\omega_k$ according to equation (\ref{eq:post-dense-omega})\;
  }
}  
 \For{$k\leftarrow 1$ \KwTo $K$}{
   Sample $\kappa_k$ according to equation (\ref{eq:post-kappa}),
   $o_k$ according to equation (\ref{eq:post-o}) 
  }
  Sample $\chi$ according to equation (\ref{eq:post-chi}),
  $\varphi$ with equation (\ref{eq:post-varphi}),
  $\pi_X$ with equation (\ref{eq:post-varpi})\;

  \For{$i\leftarrow 1$ \KwTo $p$}{
    Sample $\psi_{i,i}$ according to equation (\ref{eq:post-psi})\;
  }
}
Output $\*\Lambda$, $\*X$, $Z$, $O$\;
}
\caption{MCMC algorithm for BicMix\label{MCMC}}
\end{algorithm}

\section*{Appendix C.}
\subsection*{Data processing and comparative methods}

\subsection*{Processing the breast cancer gene expression data}

The breast cancer data set is maintained by Dana-Farber Cancer Institute at Harvard University, and is available through their R package: breastCancerNKI version 1.3.1~\citep{NKI}. We removed genes with $>10\%$ missing values. We imputed the remaining missing values using the R package {\tt impute} (version 1.36.0)~\citep{hastie-imputing:1999}. We projected the gene expression levels of each gene to the quantiles of a standard normal. There were 24,158 genes remaining in the data set after filtering.

\subsection*{Simulation comparison}
We compared BicMix to five other methods: Fabia, Plaid, CC, Bimax, and Spectral biclustering. We ran these methods using the following settings.

For Sim1, we set the number of components to the correct values, and ran each method as follows.
\begin{itemize}
\item We ran Fabia (version 2.10.2) using its default parameter settings.

\item We ran Fabia-truth using default parameter settings. We set the sparsity threshold in Fabia to the number (from $100$ quantiles of the uniform distribution over $[0.1,5]$) that produced the closest match in the recovered matrices to the number of non-zero elements in the simulated data.

\item We ran Plaid (implemented in the R package {\tt biclust}~\citep{kaiser2009biclust} version 1.0.2) using
{\tt background = TRUE} to capture the noise, maximum layers were set to $10$, number of iterations to find starting values was set to $10$, and the number of iterations to find the layers was set to $100$.

\item We ran CC (implemented in the R package {\tt biclust}~\citep{kaiser2009biclust} version 1.0.2) by setting maximum accepted score {\tt delta = 1.5} and the scaling factor {\tt alpha=1.0}.

\item We ran Bimax (implemented in the R package {\tt biclust}~\citep{kaiser2009biclust} version 1.0.2) by setting the minimum number of rows and columns to 2.

\item We ran Spectral biclustering (implemented in the R package {\tt biclust}~\citep{kaiser2009biclust} version 1.0.2) by setting the normalization method to {\tt bistochastization}, the number of eigenvalues for constructing the biclusters was set to $10$, and the minimum number of rows and columns for the biclusters were set to $2$.

\end{itemize}

For Sim2, we corrected the simulated data for the dense components by
controlling for five PCs, and we ran the methods as in Sim1, but
setting the number of components to $10$.

\subsection*{Redundancy of components}
We calculated a simple statistic to check the redundancy of the
multiple components obtained across multiple runs as follows.  For
every component across all runs, we counted the number of genes with
non-zero values, denoted as $ng$, and the number of samples with
non-zero values, denoted as $ns$ for each component. We then grouped
the components that share the same $ng$ and $ns$. For each pair of
components in the same group, we counted how many components have
non-zero values for the same genes and the same samples (i.e., the
$\ell_0$ norm). Suppose we have two components, where component $1$
has loading vector $L_1$ and factor vector of $F_1$, and component $2$
has loading vector $L_2$ and factor vector of $F_2$. We transformed
$L_1, L_2, F_1, F_2$ to binary vectors where $1$ indiciates non-zero
values and $0$ indicate zeros values. Then the number of different
genes between $L_1$ and $L_2$ is simply $ng_{dif}=\sum_{i=1}^p
(L_{1,i}-L_{2,i})^2$, and the number of different samples between
$F_1$ and $F_2$ is simply $ns_{dif}=\sum_{i=1}^n
(F_{1,i}-F_{2,i})^2$. The redundancy corresponds to the number of
pairs for which both $ng$ and $ns$ are zero.

\newpage
\section*{Appendix D.}
\subsection*{Algorithm for identifying gene co-expression networks from BicMix}

We write out the algorithm we used to build the gene co-expression networks using the fitted BicMix model. Note that the sparsity-inducing prior on the covariance matrix of the factors increases the difficulty of computing the gene-wise covariance matrix relative to the common identity matrix covariance in the prior of the factors; however, all of the elements necessary to compute an estimate of the factor covariance matrix have been explicitly quantified in the VEM algorithm already.

\begin{algorithm}[H]
\DontPrintSemicolon
\KwData{$p\times K$ loading matrix and $K \times n$ factor matrix; $\Psi$; $net\_type$, $rep$.}
\KwResult{A subset of genes }
\Begin{
  \For{$i\leftarrow 1$ \KwTo $n\_runs$}{
    \For{$k\leftarrow 1$ \KwTo $K$}{
       \small {/*use $n_{gene}[i], n_{sample}[i]$ to denote the
      number of genes and samples with non-zero values in a component at iteration $i$*/}\;
    \normalsize
      \If{$n_{genes}[5000]-n_{genes}[10000]>50$ $\&$ $n_{samples}[5000]-n_{samples}[10000]>50$}{
        discard component k
      }
    }
    \If{$net\_type$ = subset specific}{
    Add component to $A$ when non-zero factors are only in class $c$
    }
    \If{$net\_type$ = subset differential}{
      Compute Wilcoxon rank sum test for non-zero factor values across classes $c$, $d$ 
      Add component to $A$ when $p<1 \times 10^{-10}$
    } 
    Construct the covariance matrix for $\*X$ as $\*\Sigma
    \leftarrow \*{\^{XX^T}-\^X\^X^T}$ (Equations~\ref{eq:EXX} and \ref{eq:VEM-x})\;
    Calculate the variance for the residual as $\*\Psi \leftarrow $
    equation (\ref{eq:VEM-psi})\;
    Construct the precision matrix for subset $A$ as $\*\Delta^i=\(\*\Lambda^i_A \*\Sigma^i_{A,A} {\*\Lambda^i_A}^T+\*\Psi\)^{-1}$\;
    Run GeneNet~\citep{schafer_empirical_2005} on $\*\Delta^i$ to test significance of edges\;
    Store edges with probability of presence $\geq 0.8$\;
  }
  Count number of times each edge is found across all
  runs\;
  Keep edges that are found $\geq rep$ times\;
  Output the nodes and edges\;
  Draw graph using Gephi~\citep{Gephi:2009}
}
\caption{Algorithm to construct gene co-expression network\label{GGM}}
\end{algorithm}

\vskip 0.2in
\bibliography{ref}
\end{document}